\documentclass[runningheads,a4paper]{llncs}
\usepackage{amssymb,amsmath}
\usepackage[utf8]{inputenc}
\usepackage[T1]{fontenc}
\usepackage{color,geometry,verbatim}
\usepackage{pstricks,pstricks-add,pst-node}
\usepackage{algorithm,algorithmicx,algpseudocode}
\usepackage{array,epsfig,graphicx}
\usepackage[lowtilde]{url}

\newcommand{\<}{\langle}
\renewcommand{\>}{\rangle}
\DeclareSymbolFont{rsfscript}{OMS}{rsfs}{m}{n}
\DeclareSymbolFontAlphabet{\mathrsfs}{rsfscript}

\newcommand{\keywords}[1]{\par\addvspace\baselineskip
\noindent\keywordname\enspace\ignorespaces#1}

\begin{document}
\title{The \v{C}ern\'{y} conjecture for small automata: experimental report}
\author{Jakub Kowalski, Marek Szyku{\l}a}
\institute{Department of Mathematics and Computer Science, University of Wroc{\l}aw \\
\email{\{kot,msz\}@ii.uni.wroc.pl}}
\date{}
\maketitle

\begin{abstract}
We present a report from a series of experiments involving computation of the shortest reset words for automata with small number of states. We confirm that the \v{C}ern\'{y} conjecture is true for all automata with at most 11 states on 2 letters. Also some new interesting results were obtained, including the third gap in the distribution of the shortest reset words and new slowly synchronizing classes of automata.

\end{abstract}

\section{Introduction}
We deal with complete deterministic finite automata $A = \< Q,\Sigma,\delta \>$ with the state set $Q$, the input alphabet
$\Sigma$, and the transition function $\delta : \; Q \times \Sigma \to Q$. The action of $\Sigma$ on $Q$ given by $\delta$ is denoted simply by concatenation: $\delta(q,a) = qa$. This action extends naturally to the action $qw$ of words for any $w\in \Sigma^*$. If $|Qw|=1$, that is, the image of $Q$ by $w$ consists of a single state, then $w$ is called a \emph{reset} (or \emph{synchronizing}) word for $A$, and $A$ itself is called \emph{synchronizing}. The length of the shortest reset word of an automaton is called the \emph{reset length}. An automaton is \emph{irreducible synchronizing} if it is synchronizing, and the automata obtained from it by removing any letter are not synchronizing. In this paper we deal with the \emph{unlabeled model} of automata, in which the states and the letters are unlabeled. We say that two automata are isomorphic if there is a bijection between their states and a bijection between the letters, in this way isomorphic automata have the same reset length.

The \v{C}ern\'{y} conjecture states that every synchronizing automaton $A$ with $n$ states has a reset word of length $\leq (n-1)^2$. This conjecture was formulated by \v{C}ern\'{y} in 1964, and is considered the most longstanding open problem in the combinatorial theory of finite automata. So far, the conjecture has been proved only for a few special classes of automata and a general cubic upper bound has been established (see Volkov \cite{Vo2008} for general motivation and an excellent survey of the results, and Trahtman \cite{Tr2011} for a recently found new cubic bound).

The reset lengths of automata are intensively studied recently with computer aid. There are experiments (\cite{Ro2009,ST2011}) showing that the expected reset length of a random automaton with $n$ states is small and this is conjectured to be below linear bound. Another conjecture stated up by Peter Cameron during BCC conference in Exeter 2011, and in \cite{ST2011} states up that the number of synchronizing labeled automata to the number of all labeled automata tends to $1$ as $n \to \infty$. Those conjectures seem also to be true in the unlabeled model, which we discuss in the experimental section based on our results. There are dedicated computational packages implementing different algorithms to study various aspects of automata synchronization, TESTAS \cite{Tr2003} and recently developed COMPAS \cite{CR2011}.

The \v{C}ern\'{y} conjecture has been verified so far for all strongly connected automata with $k=2$ letters and $n\leq 10$ states, and $k\leq 4$ letters and $n\leq 7$ states, and $k=3$ letters and $n=8$ states \cite{Tr2006,Tr2011}. See also \cite{AGV2010} for $n=9$ states. It is sufficient to check the conjecture only for strongly connected automata. All the results were obtained with computer aid. The results of the currently performed experiments show up also interesting trends in distribution of the reset length, especially for automata with the reset length close to the upper bound $(n-1)^2$. Those are called \emph{slowly synchronizing automata}. For $6\le n\le 10$ there was found a gap in the distribution of the reset length, which is an absence of automata with the reset length in an interval just below the upper bound $(n-1)^2$ (first reported in \cite{Tr2006}). Also for $n=9$ the second gap was found, and an island between the gaps with automata with the reset length in $[n^2-3n+2,n^2-3n+4]$ \cite{AGV2010}. However these results obtained so far considered only strongly connected automata and are incomplete for $n\ge 10$, since the second gap was not reported here. Also neither of the experiments did ever reach $n=11$ states so far.

We have designed and implemented a new efficient algorithm to generate automata with small number of states and to compute the reset length for them. The method itself will be described in detail elsewhere. It allowed us to extend the known results and to verify the \v{C}ern\'{y} conjecture for all automata with $n \le 11$ states and $k=2$ letters. Also we have obtained complete distributions of the reset length for all automata $n \le 10$. For $n=11$ we report that the third gap in the distribution appears. Here we discuss the results of the experiments.

\section{Experiments}\label{sec:exp}

\subsection{The number of nonisomorphic automata}
We have computed the exact numbers, as well as the complete distribution of the reset length, of nonisomorphic automata on 2 letters up to 10 states. The results are showed in Table~\ref{tab:aut_counts_k2}. The total number of automata was known \cite{Ha1965} and we have received exactly the same numbers of automata. We present also the number of synchronizing automata, the number of strongly connected, and the number of synchronizing strongly connected automata. For the numbers of nonisomorphic strongly connected automata on 2 labeled letters see \cite{Li1971}, which are about 2 times larger than with unlabeled letters as we have considered (for example, there are 658,885 automata for $n=6$).

We can see that the fraction of synchronizing automata to all automata grows, and we may conjecture that it tends to $1$ as was conjectured for the labeled model (P. Cameron and \cite{ST2011}). This growth is more significant in strongly connected automata, the fraction of synchronizing here for $n=10$ is about $0.999$.

\begin{table}\label{tab:aut_counts_k2}
\centering
\caption{The exact numbers of nonisomorphic automata with $n$ unlabeled states on $2$ unlabeled letters in the classes of all automata, synchronizing, strongly connected, and strongly connected synchronizing. In the last column there is the fraction of the number of synchronizing automata to all automata.}
\begin{tabular}{|c|r|r|r|r|r|} \hline
$n$ & Total              & Synchronizing      & Strongly connected & S. c. and synchronizing & Synch./Total \\ \hline
2   &                  7 &                  4 &                  4 &                 2 & 0.57 \\ \hline
3   &                 74 &                 51 &                 29 &                21 & 0.69 \\ \hline
4   &              1,474 &              1,115 &                460 &               395 & 0.76 \\ \hline
5   &             41,876 &             34,265 &             10,701 &            10,180 & 0.82 \\ \hline
6   &          1,540,696 &          1,318,699 &            329,794 &           322,095 & 0.86 \\ \hline
7   &         68,343,112 &         60,477,844 &         12,310,961 &        12,194,323 & 0.88 \\ \hline
8   &      3,540,691,525 &      3,210,707,626 &        538,586,627 &       536,197,356 & 0.91 \\ \hline
9   &    209,612,916,303 &    193,589,241,468 &     26,959,384,899 &    26,904,958,363 & 0.92 \\ \hline
10  & 13,957,423,192,794 & 13,070,085,476,528 &  1,518,185,815,760 & 1,516,697,994,964 & 0.94 \\ \hline
\end{tabular}
\end{table}

\subsection{Distribution of the length of the shortest reset word}
We have computed the complete distribution of the reset length of automata on 2 letters up to 10 states, and also we have verified the \v{C}ern\'{y} conjecture for automata on 2 letters with 11 states. We confirm all the results reported in \cite{Tr2006} and \cite{AGV2010}.

\subsubsection{The exact distributions for $n \le 10$.}

The paralleled computations have been performed on 16 computers with Intel(R) Core(TM) i7-2600 CPU 3.40GHz 4 cores and 1GB of RAM. Computing the complete distribution for all automata with $n=10$ states took above 800 days of total CPU time ($\sim 13$ days of paralleled computations). Restricting to the class of strongly connected automata reduced this time to about 80 days of CPU ($\sim 2$ days of paralleled computations).

\begin{figure}\label{fig:dist10}
 \centering
 \epsfig{file=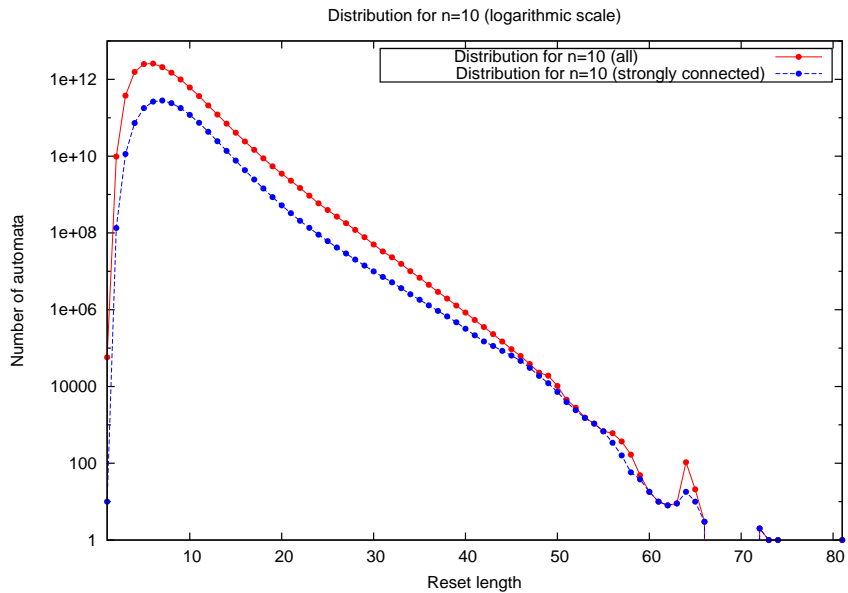, width=80mm, angle=-90}
 \caption[]{The number of nonisomorphic automata for each reset length.}
\end{figure}

Table \ref{tab:dist10} presents the obtained exact numbers of all and, separately, strongly connected nonisomorphic automata with $n=10$ states on $k=2$ letters with the longest reset lengths. The complete distribution for this case in the logarithmic scale is presented in Figure~\ref{fig:dist10}.

\begin{table}\label{tab:dist10}
\centering
\caption{The exact numbers $N(\ell)$ of all and $N_{\mathrm{sc}}(\ell)$ of strongly connected nonisomorphic automata with $10$ states on $2$ letters with the shortest reset word of length $\ell$.}
\begin{tabular}{|p{1cm}||c|c|c|c|c|c|c|c|c|c|c|c|c|c|c|c|c|c|c|c|c|c|c|c|c|c|}\hline
$\ell$                 & 56& 57& 58&59&60&61&62&63& 64&65&66&67&68&69&70&71&72&73&74&75&76&77&78&79&80&81 \\ \hline
$N(\ell)$              &607&369&168&49&18&10& 8& 9&106&21& 3& 0& 0& 0& 0& 0& 2& 1& 1& 0& 0& 0& 0& 0& 0& 1 \\ \hline
$N_{\mathrm{sc}}(\ell)$&343&160& 58&38&18&10& 8& 9&18 &10& 3& 0& 0& 0& 0& 0& 2& 1& 1& 0& 0& 0& 0& 0& 0& 1 \\ \hline
\end{tabular}
\end{table}

\begin{figure}\label{fig:dist7}
 \centering
 \epsfig{file=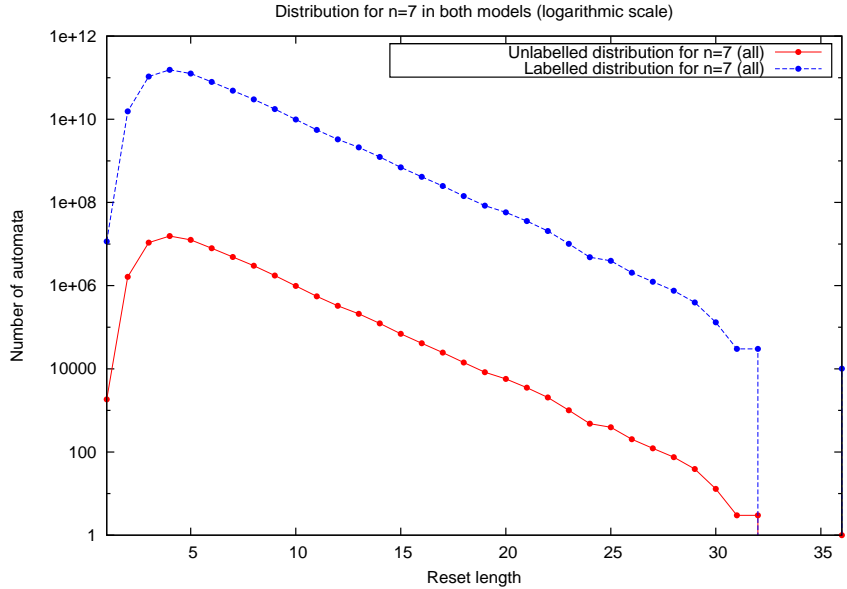, width=80mm, angle=-90}
 \caption[]{The number of automata with $7$ states on $2$ letters in the both models for each reset length.}
\end{figure}

We were also able to compute the complete distribution in the labeled model for $n \le 7$. In the labeled model the states and the letters are labeled. It may be interesting to compare regularities of both the models for $n=7$ in Figure~\ref{fig:dist7}. We see that the shape of the both the distributions is very similar. This fact gives rise to suppose that statistical properties of the distribution in the labeled model (like the fraction of synchronizing ones etc.) are also present in the unlabeled model.

\subsubsection{The distribution for $n=11$.}

We have also verified the \v{C}ern\'{y} conjecture for all automata with $n=11$ states on $2$ letters. To do this it was sufficient to restrict the tested class of automata to strongly connected. However we have not obtained the complete distribution with due to an excessive number of automata. We have computed the exact number of automata only for long reset words. By skipping synchronizing letters at the beginning we were enumerating only irreducible synchronizing automata. The number of checked strongly connected automata in this case was 79,246,008,127,339. The total CPU time in this case was above 4 years ($\sim 25$ days of paralleled computations).

Table~\ref{tab:dist11} shows the obtained numbers of all nonisomorphic automata with $n=11$ states on $k=2$ letters with the longest reset lengths. We notice that the \emph{third gap} have appeared for $n=11$ (for the report of the first gap see \cite{Tr2006}, and for the second \cite{AGV2010}). The third gap does not occur in the distribution for $n=10$. In the table there are presented some slowly synchronizing series corresponding to the reset lengths, which we describe in the next subsection.

\renewcommand{\arraystretch}{1.5}
\begin{table}\label{tab:dist11}
\centering
\caption{The numbers $N_{\mathrm{sc}}(\ell)$ of strongly connected nonisomorphic automata with $11$ states on $2$ letters with the reset length $\ell$.}
\begin{tabular}{|p{1cm}||c|c|c|c|c|c|c|c|c|c|c|c|c|c|c|c|c|c|c|c|c|c|c|c|c|c|}\hline
$\ell$                 &75&76&77&78&79&80&81&82&83&84&85&86&87&88&89&90&91&92&93&94&95&96&97&98&99&100 \\ \hline
$N_{\mathrm{sc}}(\ell)$&$\le 4$&$\le 4$&$\le 3$&0 &0 &9 &22&12&2 &1 &0 &0 &0 &0 &0 &3 &2 &1 &0 &0 &0 &0 &0 &0 &0 &1 \\ \hline
Classes                &  &  &  &--&--&  &  &$\mathrsfs{D}_n$&$\mathrsfs{W''}_n$&$\mathrsfs{\dot{B}}_n$&--&--&--&--&--&$\mathrsfs{D''}_n$&$\mathrsfs{W}_n$&$\mathrsfs{D'}_n$&--&--&--&--&--&--&--&$\mathrsfs{C}_n$ \\
                       &  &  &  &  &  &  &  &$\mathrsfs{C}^3_n$&$\mathrsfs{\dot{W''}}_n$&  &  &  &  &  &  &$\mathrsfs{B}_n$&$\mathrsfs{C}^2_n$&  &  &  &  &  &  &  &  & \\
                       &  &  &  &  &  &  &  &\ldots &  &  &  &  &  &  &  &$\mathrsfs{\dot{W}}_n$&  &  &  &  &  &  &  &  &  & \\ \hline
\end{tabular}
\end{table}

We suppose that the irregularity in the upper part of distributions -- with automata with long reset lengths -- occur also for larger numbers of states and lead to more number of gaps for larger number of states, and we may state up the gap conjecture:
\begin{conjecture}
For any natural number $g$, there exists a big enough natural number $n$ such that, in the distribution of the reset length of all nonisomorphic automata with $n$ states on 2 letters there are at least $g$ gaps.
\end{conjecture}

\subsection{Slowly synchronizing classes of automata}

By extending the results in \cite{AGV2010} to $n=11$ we have found some new slowly synchronizing classes of automata, but there is not any new in the island between the first and the second gap, all of them were found for $n=9$ states. We suppose so then that there are only these classes in the island for each $n \ge 9$:
\begin{conjecture}
In the distribution of the reset length for each number of states $n \ge 9$ and 2 letters, there is an island $[n^2-3n+2,n^2-3n+4]$ and there are exactly 4 automata $\mathrsfs{D'}_n,\mathrsfs{W}_n,\mathrsfs{D''}_n,\mathrsfs{B}_n$ if $n$ is even, and 2 more automata $\mathrsfs{C}^2_n,\mathrsfs{W'}_n$ if $n$ is odd.
\end{conjecture}

In Table~\ref{tab:dist11} we present some of the corresponding classes of automata in the distribution. Here is a survey of known slowly synchronizing classes with the longest reset length, with some new found classes.

\begin{itemize}
\item $\mathrsfs{C}_n$: $n^2-2n+1$, the \v{C}ern\'{y} automaton \cite{Ce1964}.
\item $\mathrsfs{D'}_n$: $n^2-3n+4$ \cite{AGV2010}.
\item $\mathrsfs{W}_n$: $n^2-3n+3$ \cite{AGV2010}.
\item $\mathrsfs{C}^2_n$: $n^2-3n+3$ for odd $n$, defined below.
\item $\mathrsfs{D''}_n$: $n^2-3n+2$ \cite{AGV2010}.
\item $\mathrsfs{W'}_n$: $n^2-3n+2$ \cite{AGV2010} (only mentioned there).
\item $\mathrsfs{B}_n$: $n^2-3n+2$ for odd $n$ \cite{AVZ2006,AGV2010}.
\item $\mathrsfs{\dot{W}}_n$: $n^2-3n+2$ for odd $n$, defined below.
\item $\mathrsfs{\dot{B}}_n$: $n^2-4n+7$ for odd $n$, defined below.
\item $\mathrsfs{W''}_n$: $n^2-4n+6$ \cite{AGV2010} (only mentioned there).
\item $\mathrsfs{\dot{W''}}_n$: $n^2-4n+6$ for odd $n$, defined below.
\item $\mathrsfs{D}_n$: $n^2-4n+5$ \cite{AVZ2006}.
\item $\mathrsfs{C}^3_n$: $n^2-4n+5$ for odd $n$, defined below.
\end{itemize}

\subsubsection{The automaton $\mathrsfs{W'}_n$.}
This automaton is similar to $\mathrsfs{W}_n$ but with loops instead of a cycle on the second letter:

$\delta(i,a)=\begin{cases}
i+1 &\text{if } 1 \le i \le n-1,\\
2 &\text{if } i=n;
\end{cases}\quad
\delta(i,b)=\begin{cases}
2 &\text{if } i=1,\\
1 &\text{if } i=2,\\
i &\text{if } 3 \le i \le n-2,\\
n &\text{if } i=n-1.
\end{cases}$

\subsubsection{$\mathrsfs{C}^s_n$ automata.}
We define here an generalization of the \v{C}ern\'{y} automaton $\mathrsfs{C}_n$, by parametrize the distance between the states joined by the edge on the second letter.

\begin{definition}The automaton $\mathrsfs{C}^s_n$ for $s < n$ is defined as follows:

$\delta(i,a)=\begin{cases}
i+1 &\text{if } 1 \le i \le n-1,\\
1 &\text{if } i=n;
\end{cases}\quad
\delta(i,b)=\begin{cases}
i &\text{if } 2 \le i \le n,\\
s &\text{if } i=1.
\end{cases}$
\end{definition}

Clearly $\mathrsfs{C}^1_n = \mathrsfs{C}_n$. The automata for $s > 1$ are synchronizing only for odd $n$ and their reset length is $(n-1)^2-(n-2)(s-1)$.

\subsubsection{The automaton $\mathrsfs{\dot{B}}_n$.}
This automaton is similar to $\mathrsfs{B}_n$, the first letter forms a complete circle but the second crossing edge in the second letter is placed one state later than the first crossing edge.

$\delta(i,a)=\begin{cases}
i+1 &\text{if } 1 \le i \le n-1,\\
1 &\text{if } i=n;
\end{cases}\quad
\delta(i,b)=\begin{cases}
2 &\text{if } i=1,\\
5 &\text{if } i=2,\\
i+1 &\text{if } 3 \le i \le n-1,\\
3 &\text{if } i=n-1.
\end{cases}$

The automata are synchronizing only for odd $n$ and their reset length is $n^2-4n+7$.

\subsubsection{The automata $\mathrsfs{W''}_n$ and $\mathrsfs{\dot{W''}}_n$.}
The automaton $\mathrsfs{W''}_n$:\\
$\delta(i,a)=\begin{cases}
i+1 &\text{if } 1 \le i \le n-1,\\
2 &\text{if } i=n;
\end{cases}\quad
\delta(i,b)=\begin{cases}
i &\text{if } 2 \le i \le n-3 \text{ and } i=n-1\\
1 &\text{if } i=n-2,\\
n-2 &\text{if } i=1.
\end{cases}$

The automaton $\mathrsfs{\dot{W''}}_n$:\\
$\delta(i,a)=\begin{cases}
i+1 &\text{if } 1 \le i \le n-1,\\
2 &\text{if } i=n;
\end{cases}\quad
\delta(i,b)=\begin{cases}
i &\text{if } 2 \le i \le n-3\\
2 &\text{if } i=n-1,\\
1 &\text{if } i=n-2,\\
n-2 &\text{if } i=1.
\end{cases}$

The automata are synchronizing only for odd $n$ and their reset length is $n^2-4n+6$.

\bibliography{bibliography}

\begin{thebibliography}{10}

\bibitem{AGV2010}
D.~Ananichev, V.~Gusev, and M.~Volkov.
\newblock {Slowly synchronizing automata and digraphs}.
\newblock In {\em Mathematical Foundations of Computer Science 2010}, volume
  6281 of {\em LNCS}, pages 55--65. 2010.

\bibitem{AVZ2006}
D.~Ananichev, M.~Volkov, and Y.~Zaks.
\newblock {Synchronizing automata with a letter of deficiency 2}.
\newblock In {\em Developments in Language Theory}, volume 4036 of {\em LNCS},
  pages 433--442. 2006.

\bibitem{Ce1964}
J.~{\v{C}ern\'{y}}.
\newblock {Pozn\'{a}mka k homog\'{e}nnym eksperimentom s kone\v{c}n\'{y}mi
  automatami}.
\newblock {\em {Matematicko-fyzik\'alny \v{C}asopis Slovenskej Akad\'emie
  Vied}}, 14(3):208--216, 1964.
\newblock In Slovak.

\bibitem{CR2011}
K.~Chmiel and A.~Roman.
\newblock {COMPAS - A computing package for synchronization}.
\newblock In {\em Implementation and Application of Automata}, volume 6482 of
  {\em LNCS}, pages 79--86. 2011.

\bibitem{Ha1965}
M.~Harrison.
\newblock {A census of finite automata}.
\newblock {\em Canadian journal of mathematics}, 17:100--113, 1965.

\bibitem{Li1971}
V.~A. Liskovets.
\newblock {Enumeration of non-isomorphic strongly connected automata}.
\newblock {\em {Vesci Akad. Navuk BSSR Ser. Fiz.-T\'ehn. Navuk}}, (3):26--30,
  1971.
\newblock In Russian.

\bibitem{Ro2009}
A.~Roman.
\newblock {Genetic algorithm for synchronization}.
\newblock In {\em Language and Automata Theory and Applications}, volume 5457
  of {\em LNCS}, pages 684--695. 2009.

\bibitem{ST2011}
E.~Skvortsov and E.~Tipikin.
\newblock {Experimental study of the shortest reset word of random automata}.
\newblock In {\em Implementation and Application of Automata}, volume 6807 of
  {\em LNCS}, pages 290--298. 2011.

\bibitem{Tr2003}
A.~N. Trahtman.
\newblock {A package TESTAS for checking some kinds of testability}.
\newblock In {\em Implementation and Application of Automata}, volume 2608 of
  {\em LNCS}, pages 228--232. 2003.

\bibitem{Tr2006}
A.~N. Trahtman.
\newblock {An efficient algorithm finds noticeable trends and examples
  concerning the \u{C}ern\'{y} conjecture}.
\newblock In {\em Mathematical Foundations of Computer Science}, volume 4162 of
  {\em LNCS}, pages 789--800. 2006.

\bibitem{Tr2011}
A.~N. Trahtman.
\newblock {Modifying the upper bound on the length of minimal synchronizing
  word}.
\newblock In {\em Fundamentals of Computation Theory}, volume 6914 of {\em
  LNCS}, pages 173--180. 2011.

\bibitem{Vo2008}
M.~Volkov.
\newblock Synchronizing automata and the \u{C}ern\'{y} conjecture.
\newblock In {\em Language and Automata Theory and Applications}, volume 5196
  of {\em LNCS}, pages 11--27. 2008.

\end{thebibliography}

\end{document}